\documentclass[aps,prd,twocolumn,twoside,superscriptaddress,floatfix,nofootinbib,natbib]{revtex4-1}
\usepackage{graphicx}
\usepackage{subfigure}
\newcommand{\comment}[1]{{}}
\def\beq{\begin{equation}}
\def\eeq{\end{equation}}
\def\beqn{\begin{eqnarray}}
\def\eeqn{\end{eqnarray}}

\def\cl{C_{l}}

\def\2gcm{\textrm{g cm$^{-2}$}}

\def\hires{the \emph{highRes} experiment}
\def\lores{the \emph{lowRes} experiment}

\def\H0{\ensuremath{\mathrm{H}_0}}

\def\bl{\bmm{l}}

\def\fsky{f_{\mathrm{sky}}}

\newcommand{\bmm}[1]{{\mathbf{#1}}}

\newcommand{\aap}{{Astron.~Astrophys.}}

\newcommand{\jcap}{{J.~Cosm.~Astrop.~Phys.}}
\newcommand{\araa}{{Annu.~Rev.~Astron.~Astrophys.}}

\newcommand{\procspie}{{Proc.~SPIE}}
\newcommand{\apjl}{{Astrophys.~J.~Lett.}}

\newcommand{\mnras}{{Mon.~Not.~R.~Astron.~Soc.}}

\def\ublu{\bl'}

\def\mnras{Mon. Not. R. Astron. Soc}
\def\physrep{Physics Reports}

\begin{document}
\title{Delensing the CMB with the Cosmic Infrared Background}
\author{Blake D.~Sherwin$^{*}$}
\address{Department of Physics, University of California, Berkeley, CA, 94720, USA}
\address{Miller Institute for Basic Research in Science, University of California, Berkeley, CA, 94720, USA}
\address{Berkeley Center for Cosmological Physics, LBNL and University of California, Berkeley, CA, 94720, USA}
\author{Marcel Schmittfull}
\address{Berkeley Center for Cosmological Physics, LBNL and University of California, Berkeley, CA, 94720, USA}

\begin{abstract}
As confusion with lensing B-modes begins to limit experiments that search for primordial B-mode polarization, robust methods for delensing the CMB polarization sky are becoming increasingly important. We investigate in detail the possibility of delensing the CMB with the cosmic infrared background (CIB), emission from dusty star-forming galaxies that is an excellent tracer of the CMB lensing signal, in order to improve constraints on the tensor-to-scalar ratio $r$. We find that the maps of the CIB, such as current Planck satellite maps at 545 GHz, can be used to remove more than half of the lensing B-mode power. Calculating optimal combinations of different large-scale-structure tracers for delensing, we find that co-adding CIB data and external arcminute-resolution CMB lensing reconstruction can lead to significant additional improvements in delensing performance. We investigate whether measurement uncertainty in the CIB spectra will degrade the delensing performance if no model of the CIB spectra is assumed, and instead the CIB spectra are marginalized over, when constraining $r$. We find that such uncertainty does not significantly affect B-mode surveys smaller than a few thousand degrees. Even for larger surveys it causes only a moderate reduction in CIB delensing performance, especially if the surveys have high (arcminute) resolution, which allows self-calibration of the delensing procedure. Though further work on the impact of foreground residuals is required, our overall conclusions for delensing with current CIB data are optimistic: this delensing method can tighten constraints on $r$ by a factor up to $\approx2.2$, and by a factor up to $\approx4$ when combined with external $\approx 3 \mu$K-arcmin lensing reconstruction, without requiring the modeling of CIB properties. CIB delensing is thus a promising method for the upcoming generation of CMB polarization surveys.
\comment{
As instrumental noise levels fall and confusion with lensing B-modes begins to limit experiments that search for primordial B-mode polarization, robust methods for delensing the CMB polarization sky are becoming increasingly important. We investigate in detail the possibility of delensing the CMB with the cosmic infrared background (CIB), emission from dusty star-forming galaxies that is an excellent tracer of the CMB lensing signal, in order to improve constraints on the tensor-scalar ratio $r$. We find that the maps of the CIB, such as current Planck satellite maps at 545 GHz, can be used to remove more than half of the lensing B-mode power. Calculating optimal combinations of different large-scale-structure tracers for delensing, we find that co-adding CIB data and external arcminute-resolution CMB lensing reconstruction can lead to significant additional improvements in delensing performance. We investigate whether uncertainty in the CIB spectra will degrade the delensing performance if we make minimal assumptions about their form. Though small surveys are unaffected, we find that uncertain CIB spectra can modify the delensed B-mode power in ways degenerate with $r$, thereby reducing the utility of CIB delensing for surveys larger than a few thousand square degrees. However, when simultaneously constraining $r$ and the true CIB spectra, we find that the delensing procedure can be self-calibrated by measurements of the delensed B-mode power spectrum on smaller scales $\ell = 100-450$, thereby preserving the viability of CIB delensing even for large surveys (though dust foregrounds may be a limiting factor for such surveys). We conclude that delensing with CIB maps can tighten constraints on $r$ by a factor of more than $2$, and by a factor close to $4$ when combined with external $\approx 3 \mu$K-arcmin lensing reconstruction, without sensitivity to uncertainties in modeling the CIB. CIB delensing is thus a promising method for the upcoming generation of CMB polarization surveys.}
\end{abstract}
\maketitle
\section{Introduction}
Many inflationary models make a key prediction:~the production of a significant background of gravitational waves \citep{grishuk, staro, rubakov, fabbri, abbott,lyth97}.~A detection of this stochastic gravitational wave background would therefore provide particularly strong evidence for the inflationary paradigm, alongside the observed super-horizon correlation, near scale-invariance and Gaussianity of the scalar density fluctuations (e.g., \cite{planckinflation} and references therein). Such a detection would also give significant insights into the physics of inflation. In particular, the ratio of gravitational wave or tensor power to scalar density power $r$ gives a direct probe of the energy scale at which inflation occurred. \footnotetext{email:~sherwin@berkeley.edu}

The currently most promising method to measure this inflationary gravitational wave background is through its effects on the cosmic microwave background (CMB). However, tensor modes can be difficult to constrain accurately via the CMB temperature alone, because scalar modes also contribute to the temperature, which leads to cosmic variance and degeneracies. Unlike in temperature, the CMB sky contains two degrees of freedom in polarization (a magnitude of polarization and a direction); because of this, one can construct a ``null" linear combination from the two degrees of freedom which has no response to scalar perturbations at linear order, but still is sourced by inflationary gravitational waves. This combination, known as the CMB B-mode polarization, therefore allows a constraint on primordial gravitational waves that is robust and free of leading-order scalar cosmic variance \citep{seljak97,kamionkowski97,seljakzaldarriaga97}. 

Over the past year, the observational status of primordial B-mode polarization has been a subject of active debate. While the BICEP2 collaboration reported a measurement of B-mode polarization on large angular scales \citep{bicep2}, subsequent analyses based on Planck polarization measurements suggested that this signal could be sourced by foreground emission from polarized galactic dust \citep{mortonson14, flauger14, planckdust}. A recently-released cross-correlation analysis involving data from the BICEP2, Keck Array and Planck experiments \citep{bicepCross} has confirmed that a significant fraction of the signal seen in BICEP2 was of galactic origin, and has thus replaced any evidence for inflationary gravitational waves with a new upper limit of $r<0.12$ at 95\% confidence. For simplicity, and motivated by this new result, we work under the default assumption that $r=0$ (though cosmic variance from tensor modes could straightforwardly be incorporated into our analysis if future measurements required this). While foreground polarization appears to be a significant obstacle when analyzing only one frequency channel, component separation with multifrequency data is a promising path towards removing much of the foreground contamination. \comment{Over the next decade, increasingly sensitive telescopes will survey the CMB polarization sky at multiple frequencies (e.g., BICEP2, Keck Array, BICEP3, SPIDER, ABS, EBEX, ACTPol, SPTpol, POLARBEAR-II, Simons Array, LiteBIRD, CoRE, PIXIE, CMB-S4), giving imeasurements of foreground-cleaned B-mode polarization on large angular scales.}

Even in the absence of foregrounds, primordial gravitational waves are not the only source of B-mode polarization. A well-understood, recently observed \citep{hanson,pbcross,pbauto,pbauto2,bicep2,actpollens,story,bicepCross} source of B-mode polarization is the gravitational lensing of the CMB, the deflection of CMB photons by the distribution of large-scale structure along the photons path. By remapping the CMB sky, gravitational lensing converts some of the primordial E-mode polarization (the polarization component that can be sourced by scalars) into B-mode polarization \citep{zalsel}. The lensing B-mode power spectrum is constant at low multipoles, so that the lensing B-modes induce variance and act as a source of noise similar in effect to instrumental white noise at a level of $\sim 4.9\mu$K-arcmin. The Keck Array / BICEP2 Collaboration recently reported the measurement of CMB maps that, for the first time, have noise levels below this value ($3.4\mu$K-arcmin, \cite{kecknew}). Lensing B-mode noise is thus beginning to limit even current B-mode surveys.  Hence, for an improved constraint on $r$ (or a better characterization of the signal if it is non-zero), delensing methods are rapidly becoming essential \citep{knoxsong, kesden, seljakhirata}. These methods involve subtracting off an estimate of the lensing B-mode map in order to reduce the lensing noise. 

To construct an estimate of the lensing B-mode and delens, one requires both a measurement of the E-mode polarization and an estimate of the lensing which remapped the primordial CMB sky. Internal methods to delens involve reconstructing the lensing field (using statistical methods such as a quadratic estimator, \cite{huokamoto},  or a full maximum likelihood approach, \cite{hirataseljak, seljakhirata}) with the same CMB dataset that is used to constrain the primordial B-modes. However, very low noise levels are required for the best internal delensing performance. Furthermore, much experimental progress continues to be made with CMB experiments with large beams ($\gg 10'$), which cannot reconstruct CMB lensing maps that are sufficiently signal-dominated for efficient delensing. With such experiments reaching low noise levels, delensing which makes use of external datasets to estimate the lensing field may be promising. 

The cosmic infrared background (CIB) is diffuse far-infrared radiation originating in mostly unresolved dusty star-forming galaxies, where it is emitted by the UV-heated dust enshrouding young stars \citep{hauser}. In a number of studies, the CIB was found to have a remarkably high correlation ($\approx 80\%$) with CMB lensing (e.g., \cite{planckciblensing, holder}), confirming earlier predictions based on the similar high-redshift origin of both signals \citep{song}. This high correlation makes the CIB an ideal proxy for CMB lensing. In this paper, we will investigate the use of maps of the Cosmic Infrared Background to delens the B-mode sky.

The use of external large-scale structure tracers for delensing was investigated in \cite{smithdelensing}, focusing on tracers from surveys with well-known redshift origin. For such tracers, the authors concluded that external delensing was not a very promising technique, as futuristic surveys that mapped out all the structure to redshifts beyond $z>2-3$ were required to achieve good delensing performance. Unlike the surveys considered, CIB maps are well-suited for delensing, because even currently available CIB maps trace the underlying mass distribution over a broad range of high redshifts and thus are excellent lensing tracers \citep{vieroa,vierob, bethermin}. In a paper discussing how well delensing methods can improve constraints on the tensor tilt $n_T$ (which appeared as this paper was being completed), \cite{simard}, the authors included forecasts for a constant-correlation CIB channel. Our work differs because it focuses on constraints on the tensor amplitude $r$ rather than tensor tilt and assumes $r$ is significantly less than 0.2. We also discuss additional aspects of CIB delensing, such as combinations of datasets, and, in particular, the effects of uncertainties in the CIB cross- and auto-spectra.

After briefly reviewing delensing in general, we discuss methods for CIB delensing and for calculating the resulting reduction in lensing noise in section 2. We then apply this formalism to models of the Planck CIB data in section 3 in order to forecast how much detection errors on $r$ can be improved by CIB delensing. In section 3, we also discuss combinations of external delensing datasets: first, the optimal combination of CIB frequencies, and second, the combination of the CIB with external estimates of lensing from high-resolution CMB experiments. In section 4, we investigate the limitations of real data, focusing in particular on the problem of uncertainty in the CIB cross- and auto-spectra, and investigating to what extent CIB delensing can be self-calibrated without assuming models for the CIB spectra. We conclude in section 5. In this paper we assume a flat $\Lambda$CDM cosmology with $\{\Omega_\mathrm{b} h^2, \Omega_\mathrm{c} h^2, H_0, 10^9 A_\mathrm{s}, n_\mathrm{s}, \tau\} = 
\{0.02226, 0.1193, 67.27, 2.130, 0.9653, 0.063\}$,~which is very close to the TT,TE,EE+lowP+lensing best-fit base $\Lambda$CDM model from \cite{planckparams}.

\section{Lensing, Delensing and Correlation}
\subsection{Preliminaries: Noise from Lensing B-modes}
As CMB photons travel through the universe, they are gravitationally lensed by large-scale structure. This lensing leads to a remapping of the CMB polarization anisotropies from their unlensed position $\mathbf{n}$ on the sky:
\beq
Q(\hat{\mathbf{n}}) = Q_{\mathrm{unlensed}}(\hat{\mathbf{n}}+\mathbf{d});~~
U(\hat{\mathbf{n}}) = U_{\mathrm{unlensed}}(\hat{\mathbf{n}}+\mathbf{d})
\eeq
with an analogous remapping in temperature. Here $Q$ and $U$ are the CMB polarization Stokes parameters, and $\mathbf{d}$ is the lensing deflection field. As the curl-like part of the deflection is many orders of magnitude smaller than the gradient-like part \citep{hirataseljak}, the lensing can be equivalently described using the convergence $\kappa = -\frac{1}{2} \nabla \cdot \mathbf{d}$, which can be shown to equal a projection of the matter density out to high redshift \citep{lewischallinor}. The convergence is also simply related to the often-used lensing potential $\phi$ by $\kappa =-\frac{1}{2}\nabla^2 \phi$.

Writing the polarization not in terms of the experimentally measured $Q$ and $U$ parameters but in terms of even parity E and odd parity B-modes, we find that lensing induces a conversion of a pure E mode pattern into B-modes, with the lensing B-mode given in harmonic space by (to leading order in $\kappa$):
\beq
B^{\mathrm{lens}}(\bl) =  \int \frac{d^2 \bl'}{(2 \pi)^2} W(\bl,\bl') E(\ublu) \kappa(\bl - \bl')
\eeq
for
\beq
W(\bl,\bl') = \frac{2 \bl' \cdot (\bl-\bl')}{|\bl-\bl'|^2} \sin(2\varphi_{\bl,\bl'}),
\eeq
where $E$ is the unlensed E-mode, $l, \bl$ are multipoles written in the flat sky approximation, and $\varphi_{\bl,\bl'}$ is the angle between $\bl$ and $\bl'$. The flat-sky approximation used here gives sufficient accuracy for our simple forecast calculations and $l$-range, though we also verify our numerical results using the corresponding full-sky formulae \citep{huCMB}. Keeping only the lowest order term in $\kappa$ has also been found to be an excellent approximation on the scales relevant for delensing \citep{smithdelensing,challewis}.

From this expression, we can calculate the power spectrum
\beq
\langle B^{\mathrm{lens}}(\bl) B^{\mathrm{lens}^*}(\tilde{\bl}) \rangle \equiv (2\pi)^2 \delta^D(\bl - \tilde{\bl}) C^{BB,\mathrm{lens}}_l 
\eeq
to give
\beq
C_l^{BB,\mathrm{lens}}  = \int \frac{d^2 \bl'}{(2 \pi)^2} W^2 (\bl,\bl')C^{EE}_{l'} C^{\kappa \kappa}_{|\bl-\bl'|}
\eeq
This lensing B-mode power is effectively constant on large scales $l<100$ \citep{smithdelensing}.
 
The full (beam-deconvolved) B-mode power spectrum is given by
\beq
C_l^{BB,\mathrm{full}} = C_l^{BB,r} + C_l^{BB,\mathrm{lens}} + N_l^{BB}
\eeq
where ${C}_l^{BB,r}$ is the contribution of tensor modes to the B-mode power spectrum (assumed zero by default) and $N^{BB}_l = N^{EE}_l $ is the instrumental noise power
\beq
N_l^{BB} = \left({\Delta_P}/{T_{\mathrm{CMB}}}\right)^2  e^{ {l^2 \theta_\mathrm{FWHM}^2}/({8 \ln 2})}
\eeq
with $\Delta_P$ the noise level, here written in $\mu$K-radians (though typically specified in $\mu$K-arcmin), and $\theta_\mathrm{FWHM}$ the full width at half maximum of the beam in radians.
Assuming the realization of the lensing B-modes on the sky is unknown, the presence of lensing B-mode power will contribute to the error on a measurement of the B-mode power spectrum:
\beq
\sigma( C_{l}^{BB,\mathrm{full}} )= \sqrt{ \frac{2}{(2 {l}+1) \fsky }} \left( C_{{l}}^{BB,\mathrm{lens}}+N_{{l}}^{BB} \right)
\eeq
where $\fsky$ gives the sky fraction observed by the survey. While non-Gaussian covariance between different B-multipoles is present because fluctuations in lensing and E-modes simultaneously affect different B-mode scales \citep{smithbold,gayoung,smithbnew,chao,benoit}, this covariance is subdominant on the large scales that constrain $r$. Despite this, we include this covariance in our analysis in later sections of this paper. The error on the B-mode power determines the error on a constraint of the tensor-scalar ratio $r$ (assuming the ``detection'' null hypothesis of no tensor modes in the error calculation):
\beqn
&\sigma(r)&= \left[\sum_l \frac{\left(\frac{\partial {C}_l^{BB,r}}{\partial r}\right)^2}{\sigma^2( C_{l}^{BB,\mathrm{full}} )} \right]^{-\frac{1}{2}}  \\ &\approx&  \left[ \frac{\sum_l (2 {l}+1) \fsky \left(\frac{\partial {C}_l^{BB,r}}{\partial r}\right)^2}{2} \right]^{-\frac{1}{2}}\langle C_l^{BB,\mathrm{lens}} +N^{BB}_l \rangle_{l<100}\nonumber
\eeqn
In the second line we have assumed that both the lensing B-mode and noise power spectra are constant, i.e., white-noise-like, at $l<100$  \citep{smithdelensing}, where nearly all the constraining power on $r$ arises (note that the derivative with respect to $r$ is simple since $r$ merely normalizes a fixed tensor spectrum). The lensing-induced error affects any measurement of the B-mode power spectrum and therefore any constraint on $r$, and persists even as instrumental noise is reduced to zero. Delensing algorithms are therefore needed to remove the lensing B-mode and thus reduce the errors when the instrumental noise approaches or falls below the level of $\cl^{BB,\mathrm{lens}}$, which corresponds to $\approx 4.9\mu$K-arcmin in polarization.

\subsection{Delensing with Large-Scale Structure: Methods and Expected Performance}
We now consider delensing with a cosmic infrared background map $I$, which traces the underlying matter distribution and which we will attempt to use as a proxy for $\kappa$. (Our analysis is general and thus $I$ could be any tracer of large-scale structure.) We construct an estimate of the lensing B-mode using this field, using a weighting $f(\bl,\bl')$ (which we will determine):
\beq
\hat{B}^{\mathrm{lens}}(\bl) = \int \frac{d^2 \bl'}{(2 \pi)^2} W(\bl,\bl') f(\bl,\bl') E^N(\ublu) I(\bl - \bl')
\eeq
where $E^N$ is the noisy, observed E-mode; to lowest order, the fact that it is lensed can be neglected. Though we distinguish $E$ and $E^N$, we will always take $I$ to contain all the components of the observed CIB map (e.g., shot noise, foregrounds, instrumental noise if relevant).

We then subtract this estimate off from the measured B-mode in order to delens. The residual lensing B-mode is given by 
\beqn
B^\mathrm{res}(\bl) &=&  B^\mathrm{lens}(\bl) - \hat{B}^\mathrm{lens}(\bl) =  \int \frac{d^2 \bl'}{(2 \pi)^2} W(\bl,\bl') \times \nonumber \\ &&\left( E(\ublu) \kappa(\bl - \bl' ) -  f(\bl,\bl') E^N(\ublu) I(\bl - \bl' ) \right) 
\eeqn

We now wish to determine $f$ such that the residual lensing B-mode power, and hence the lensing B-mode noise, is minimized. Evaluating the lensing B-mode power spectrum after the delensing term has been subtracted gives:
\beqn
C_{{l}}^{BB,\mathrm{res}} &=&   \int \frac{d^2 \bl'}{(2 \pi)^2} W^2 (\bl,\bl') 
 [ C^{EE}_{l'} C^{\kappa \kappa}_{|\bl-\bl'|} \nonumber \\ && -(f(\bl,\bl')+f^*(\bl,\bl')) C^{EE}_{l'} C^{\kappa I}_{|\bl-\bl'|} \\ \nonumber && + f^*(\bl,\bl') f(\bl,\bl') (C^{EE}_{{l'}}+ N^{EE}_{{l'}}) C^{II}_{|\bl-\bl'|} ] \eeqn
Taking the derivative with respect to $f$, we obtain the form of $f$ which minimizes the residual lensing B-mode:
\beq
f(\bl,\bl') = \left(\frac{C^{EE}_{{l'}}}{C^{EE}_{{l'}}+N^{EE}_{{l'}}}\right)  \frac{C^{\kappa I}_{|\bl-\bl'|}}{C^{II}_{|\bl-\bl'|} } 
\eeq
We note that for signal-dominated $E$-modes, this weighting corresponds simply to a normalization of the $I$ field by ${C^{\kappa I}_{l}/C^{II}_{l}}$. Additionally, noisy $E$-modes are down-weighted.   We also note that the same weight $f$ could be obtained by treating the part of the CIB that is uncorrelated with lensing as effective noise in the standard calculation of internal delensing methods as in \cite{smithdelensing} and \cite{simard}. However, our explicit approach has advantages when we consider uncertainty in the CIB spectra in later sections of this paper.

Using this (perturbatively) optimal weight $f$, we obtain the following residual lensing B-mode power after delensing:
\beqn
C_{{l}}^{BB,\mathrm{res}} &=&   \int \frac{d^2 \bl'}{(2 \pi)^2}  W^2 (\bl,\bl') 
 C^{EE}_{l'} C^{\kappa \kappa}_{|\bl-\bl'|}  \\
&\times&  \left[1 - \left(\frac{C^{EE}_{{l'}} }{C^{EE}_{{l'}}+N^{EE}_{{l'}}}\right) \rho^2_{|\bl-\bl'|} \right] \nonumber
\eeqn
where $\rho$ is the correlation coefficient of CMB lensing and the CIB:
\beq
\rho_l= \frac{\cl^{\kappa I}}{\sqrt{\cl^{\kappa \kappa} \cl^{I I}}}
\eeq
The full delensed B-mode power spectrum then becomes (from Eq.~6):
\beq
C_l^{BB,\mathrm{del}} = C_l^{BB,r} + C_l^{BB,\mathrm{res}} + N_l^{BB}.
\eeq
When the E-mode is signal dominated for all relevant $l'$ used in the sum in Eq.~14, the delensing can be simply described by the following reduction of the lensing signal in the map:
\beq
C^{\kappa \kappa}_l \rightarrow (1-\rho_l^2)C_l^{\kappa \kappa}.
\eeq
If the correlation coefficient is constant with $l$, the lensing B-mode power and the resulting lensing noise are then similarly reduced by a factor $ (1-\rho^2)$.

\section{Delensing with the CIB}
\subsection{Modeling the CIB-Lensing Correlation for Forecasting Delensing Performance}
To estimate the delensing performance and the reduction in noise for B-mode searches, we must know the residual B-mode power; to determine the residual B-mode power with Eq.~14, we must in turn have an estimate for the correlation coefficient of the CIB with CMB lensing, $\rho_l$. To calculate $\rho_l$ in this work, we will use the CIB halo model of \cite{planckauto} to estimate CIB lensing-cross- and auto-power spectra; this model is a good fit to Planck measurements of these spectra \cite{planckciblensing, planckauto}. As the CIB power spectra in the default model do not include the shot noise contributions that remain in the measured spectra, we add shot noise terms for both bright CIB and radio sources; we obtain the relevant amplitudes (which match the measured spectra) from the models used in \cite{planckauto}. However, in this section we do not include instrumental noise (or any additional noise due to component separation). On the scales relevant for delensing, this is an excellent approximation for the higher frequency Planck channels (545, 857 GHz), but not necessarily the lower frequency channels (353 GHz, 217 GHz), which are only included for comparison. We will discuss the impact of uncertainties of CIB spectra in section 4.

The correlation coefficients for the different frequencies are shown in Fig.~1. As expected from prior work \citep{planckciblensing}, we find a high degree of correlation at low multipoles, peaking above 0.8 for all the frequencies. This again illustrates that the CIB is an excellent tracer of lensing. The correlation coefficients do not appear constant with scale, unlike shown in Figure 13 of \cite{planckciblensing}; in part, this is because we include the shot noise contribution to the CIB power spectra, though even without the inclusion of shot noise the correlation in the model we use falls somewhat with $l$. We will assume that only CIB scales $l>60$ can be used to delens; however, the results are not very sensitive to the exact value of the low-$l$ cutoff, as most of the lensing B-modes arise from higher-$l$ lens scales (see Fig.~1).

\begin{figure}[h!]
\includegraphics[width=\columnwidth]{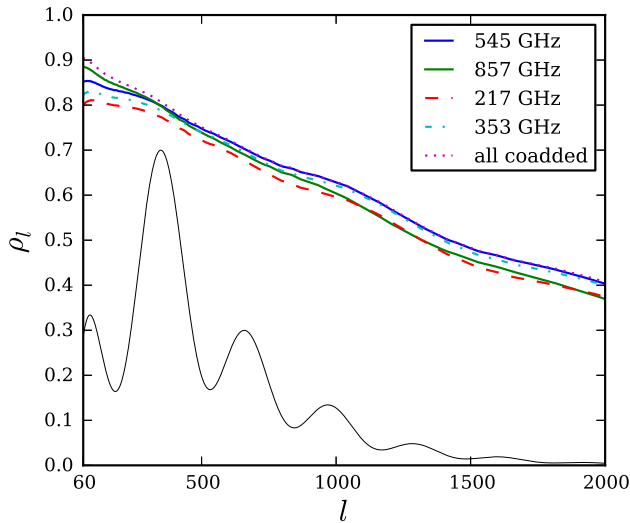}
\caption{Correlation coefficients for the different Planck CIB frequencies (545, 857, 217, 353 GHz) with CMB lensing, using the best-fit model from \cite{planckauto}. We also show the correlation coefficient for the co-added linear combination of these CIB frequencies. The contribution of shot noise to the correlation coefficient is included (at a level which matches the measured CIB spectra). We neglect instrumental noise (or any additional noise due to component separation), which is an excellent approximation for the high frequency Planck channels (545, 857 GHz), but not necessarily the lower frequency channels; these are thus only included for comparison. Finally, plotted beneath the correlation curves, the black solid line shows $\langle C_{l}^{\kappa \kappa} \times \partial C_{l'}^{BB}/\partial C_l^{\kappa \kappa}\rangle_{l'<100}$ (in arbitrary units), i.e.~how much each scale $l$ in the lensing contributes to the mean B-mode power on scales $l'<100$, and thus which scales are most important to delens.}
\end{figure}

We now consider to what extent foregrounds, in particular galactic dust emission, will reduce the correlation from the models described above and degrade the delensing performance. To determine this, we consider the effect of foregrounds on the CIB maps' cross-correlations with CMB lensing as well as on their power spectra. Galactic foregrounds are not expected to be correlated with the true lensing maps (so that the cross-power measured in \cite{planckciblensing} should be accurate everywhere), but they will contribute to the CIB auto-power spectra on large scales. However, on $\approx2240 \ \mathrm{deg}^2$ of sky, the Planck Collaboration \cite{planckauto} have demonstrated that galactic dust can be removed by subtracting dust maps constructed from observations of galactic HI emission (21cm emission from neutral hydrogen, \cite{kalberla}) combined with Planck HFI and IRAS observations; after subtraction, the dust contributions to the auto-spectra are small on angular scales above $l = 150$. While the Planck CIB analyses that demonstrate galactic dust subtraction rely on some of the cleanest areas of sky, most upcoming B-mode searches will either focus on these same low-dust regions at high galactic latitude where foreground contamination is smallest, or will have enhanced multifrequency coverage that will allow improved dust modeling and subtraction. Even if large scales $l<150$ are assumed completely unusable, the reduction in delensing performance when changing from $l_\mathrm{min}=60$ to $l_\mathrm{min}=150$ is only $\approx12\%$, because smaller scales $l=200-500$, which are less affected by dust, are much more important for delensing (as can be seen in Fig.~1). Though for the largest surveys our forecasts may be somewhat optimistic, our correlation model should thus be sufficiently accurate to make realistic forecasts of delensing performance for most surveys. In our analysis we have assumed that the E and B CMB polarization maps have been cleaned with multifrequency data so the levels of residual galactic dust and CIB contamination in the CMB maps are negligible, though we revisit this in section 4, along with other aspects of potential foreground contamination.

\comment{As the delensing performance is set by the CIB spectra which are used to construct the models we use for the correlation coefficient, this indicates that residual dust contamination should only reduce the delensing effectiveness by a small amount compared to the foreground-free performance. }

\subsection{Delensing Performance: Individual Frequencies}
With this model for the correlation coefficients, we can characterize the delensing performance for CIB maps at different frequencies. We calculate the delensed residual B-mode power for the different Planck CIB channels from Eq.~14, at first setting the instrumental noise in the CMB maps to zero. The residual lensing B-mode spectra, after delensing in this idealized case, are shown in Fig.~2; we find that for all Planck frequencies CIB delensing can remove a substantial part of the lensing-B mode power. We quantify this by measuring the change in the mean $\cl^{BB,\mathrm{lens}}$ after delensing, averaged over large scales $l<100$, where most of the constraining power for primordial B-mode polarization lies. We obtain the following results: after delensing noise-free CMB maps using the 857 GHz,  545 GHz, 353 GHz, and 217 GHz CIB frequencies, the mean low-ell $\cl^{BB,\mathrm{lens}}$ is reduced by a factor 0.46, 0.45, 0.47, 0.49 respectively (though the 217 and 353 GHz delensing performance will be degraded when noise is included.) With our knowledge of the residual B-mode power, we can now define an improvement factor in the measurement error of $r$, $\sigma(r)$, as in \cite{smithdelensing}: 
\beq
\alpha \equiv \sigma_0(r)/\sigma_\mathrm{delensed}(r)
\eeq
where $\sigma(r)$ depends on the B-mode power as in Eq.~9, $\sigma_0$ indicates the error before delensing, and $\sigma_\mathrm{delensed}$ is the error after delensing (i.e., determined by $C_l^{BB,\mathrm{res}}$ rather than $C_l^{BB,\mathrm{lens}}$). When the instrumental noise in Eq.~9 is neglected, this reduces to the simple expression $ \alpha = \langle C_l^{BB,\mathrm{lens}} \rangle_{l<100}/ \langle C_l^{BB,\mathrm{res}}\rangle_{l<100}$ because both the lensed and residual B-mode power are flat at $l<100$ (see Fig.~2); we show the results for this expression (which is just the inverse of the reduction factors quoted earlier) in Table 1.
\begin{table}
\begin{center}
  \begin{tabular}{| c | c |}
    \hline
    Frequency & Error Improvement Factor $\alpha$\\ \hline 
    857 GHz & 2.2  \\
    545 GHz & 2.2  \\ 
    353 GHz & 2.1 \\
    217 GHz & 2.0  \\
    545 + 857 GHz & 2.3  \\
    all frequencies coadded & 2.3  \\
    \hline
 
  \end{tabular}
\end{center}
     \caption{Ideal improvement factors on errors on a detection of $r$ achievable with CIB delensing using four Planck frequency channels, assuming zero instrumental noise in both the CMB and the CIB maps (a good approximation for 545 and 857 GHz, but not for the lower frequencies).}
\end{table}

Going beyond the ideal noise-free limit, we now examine the improvement in measurement errors on $r$ as a function of instrumental noise level of a CMB polarization telescope. With non-zero experimental noise, the improvement factor becomes:
\beq
\alpha = \frac{ \langle C_l^{BB,\mathrm{lens}} +N^{BB}_l[\Delta_P] \rangle_{l<100}}{\langle C_l^{BB,\mathrm{res}}[\Delta_P]+N^{BB}_l[\Delta_P] \rangle_{l<100}}
\eeq
where the residual B-mode power is calculated from Eq.~14 as before, now including the effects of instrumental noise in the calculation.

Since the noise power spectrum after beam deconvolution enters into this calculation, the impact of noise depends on the instrumental beam size (as seen in Eq.~7). We will thus consider two reference experiments in our analysis. The first reference experiment, which we will designate the \emph{highRes} experiment, has high resolution -- a 1.4 arcmin FWHM Gaussian beam -- and is thus capable of efficient internal delensing when noise levels are low. This experiment should be taken as a proxy for ground-based telescopes such as ACTPol, POLARBEAR-I/II, SPTpol, and their successors \cite{niemack,pb2,sptpol}, as well as high-resolution satellites such as COrE or PRISM \citep{core, prism}. The second reference experiment, which we term the \emph{lowRes} experiment, is a lower angular resolution CMB telescope targeted towards measuring large-scale CMB polarization. We take it to have a 30 arcmin FWHM Gaussian beam; this resolution is too low for efficient internal delensing. The \emph{lowRes} experiment is intended to represent CMB telescopes such as Keck Array, BICEP3 or LiteBIRD \cite{kernasovskiy,bicep3,litebird}. For \lores, we assume that the measured E-mode can be co-added with Planck maps of the E-mode polarization on smaller scales (for which we assume 60$\mu$K-arcmin in polarization and a 7-arcmin beam, as in \cite{pearson}).

For both experiments, we show the improvement factor $\alpha$ for delensing with the 545 GHz CIB frequency in Fig.~3, as a function of polarization noise level. Other CIB frequencies give nearly the same results, as their correlation coefficients with lensing are very similar. For \hires, we see that $\alpha$ becomes greater than 1.5 as noise levels fall below 4$\mu$K-arcmin and reaches $\approx 2.2$ as the experimental noise approaches zero. As a CIB-delensed survey gives the same constraints as an $\alpha^2$ times larger survey that has not been delensed (neglecting windowing effects), these improvement factors can correspond to significant enhancements in experimental power. For \lores, the delensing performance, though similar, is slightly reduced for intermediate noise levels. This is due to differences in the E-mode map: while for \hires, the E-mode map is signal-dominated on all relevant scales (i.e., the results do not change visibly if we set the noise in the E-mode map to zero), the \emph{lowRes} E-mode is not signal-dominated everywhere, which results in a small reduction in delensing performance.

For \hires, we also show the improvement factor $\alpha$ for internal delensing, i.e.~for delensing with an estimate of the lensing convergence reconstructed from CMB data. We calculate the expected noise level of CMB lensing reconstruction for the minimum variance combination of optimal quadratic estimators as in \cite{huokamoto} (assuming scales of $l=100-3000$ and $l=100-4000$ are used in temperature and polarization reconstruction respectively); we hence obtain a correlation coefficient for the reconstructed maps, which allows us to calculate the improvement factor $\alpha$ as before and compare with CIB delensing.

In Fig.~3, we see that for noise levels above \comment{TODO: change numbers if end up using Planck15 cosmology.}$\approx 5-6 \mu$K-arcmin, CIB delensing performs slightly better  (though $\alpha$ is not large); for much lower noise levels, with a small beam such that lensing reconstruction is possible, internal lensing reconstruction performs better than the CIB at delensing \hires. The use of an iterative procedure for internal reconstruction will not greatly modify the reconstruction curves at the noise levels shown in Fig.~3 \citep{smithdelensing}, though it will cause large increases to $\alpha$ when the noise is below $\approx 2 \mu$K-arcmin. It is encouraging that for moderate noise levels $>2 \mu$K-arcmin, internal delensing and CIB delensing (with current CIB data) are fairly similar in performance. Furthermore, for future \emph{lowRes}-like experiments without the necessary angular resolution to perform lensing reconstruction, CIB delensing will allow an improvement at low noise levels.

\begin{figure}[h!]
\includegraphics[width=\columnwidth]{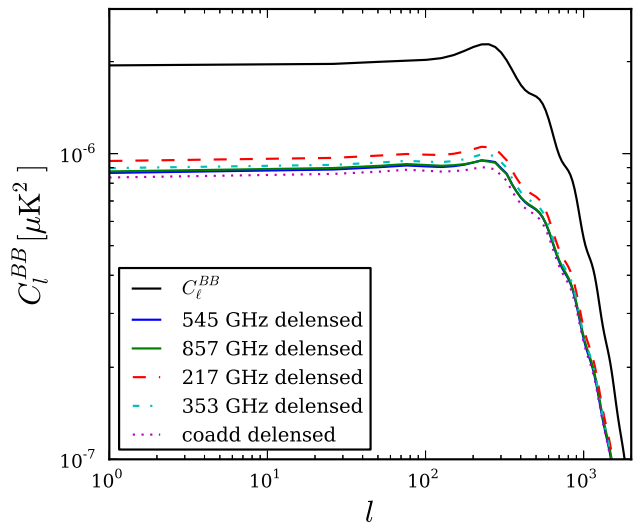}
\caption{B-mode power spectra after delensing with CIB maps at different frequencies (and with a linear combination of maps at all frequencies). The upper solid line shows the original lensing B-mode power spectrum. We here assume that the CMB maps have negligible instrumental noise on all relevant scales. As in Fig.~1, we also neglect instrumental and component separation noise in the CIB maps (which is a good approximation for the Planck 545 and 857 GHz maps).}
\end{figure}

\begin{figure}[h!]
\subfigure{\includegraphics[width=\columnwidth]{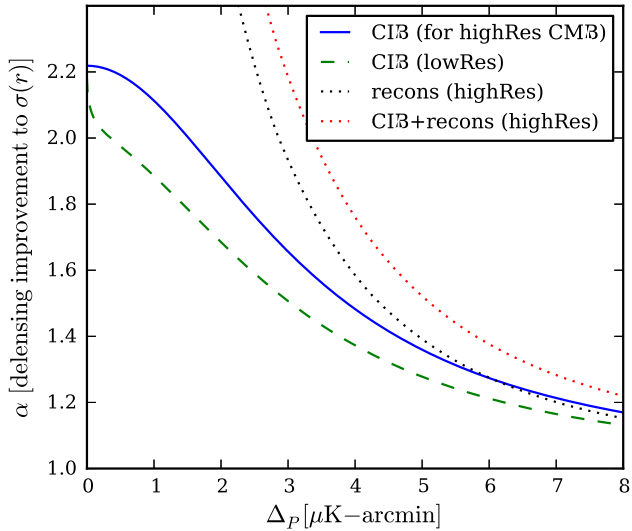}}
\caption{Improvement factors in errors on a detection of $r$ when delensing with the 545 GHz CIB channel, as a function of the polarization noise level of the experiment. The blue solid line shows CIB delensing performance for an experiment with a $1.4$ arcmin beam (our \emph{highRes} reference experiment). The green dashed line shows results for an experiment with a 30 arcmin beam (\lores). For this experiment, we assume the E-mode map used for delensing is obtained in combination with E-mode maps from Planck.  It can be seen that for low noise levels, CIB delensing can improve constraints on $r$ by a factor of up to $\approx2.2$. \comment{Our forecasts depend somewhat on whether the parameters used correspond to a Planck or WMAP9 cosmology, also shown with a thin dot-dashed line; assuming the WMAP9 lensing power spectrum leads to higher forecasted delensing performance).} For comparison, we also plot the improvement when delensing \hires ~using internally reconstructed (minimum variance quadratic estimator) lensing maps (this requires high resolution). $\alpha$ is shown both for lensing reconstruction alone (dotted black line), and for lens reconstruction coadded with the 545 GHz CIB (dotted red line).
}
\end{figure}

\subsection{Combining Different CIB Frequencies}
We now investigate how well we can delens using combinations of large-scale-structure datasets, turning first to combinations of different CIB channels. Writing our combined delensing tracer $I$ as a linear combination $I= \sum c_i I_i$, where the fields $I_i$ are different CIB frequencies (or other large-scale structure tracers), we must determine the coefficients $c_i$ that allow us to delens optimally. We note that in Eq.~14, these coefficients only enter into the correlation coefficient of $I$ with lensing, $\rho_l$; hence, if we seek to maximize the delensing performance, we must maximize $\rho_l$ with respect to the $c_i$ coefficients. In the appendix, we solve for the coefficients $c_i$ that maximize the correlation. We find that for the optimal linear combination $I= \sum c_i I_i$,
\beq
c_i = \sum_j (\rho^{-1})_{ij} \rho_{ j \kappa } \sqrt{\frac{\cl^{\kappa \kappa}}{\cl^{I_i I_i}  }} 
\eeq
where $\rho^{i \kappa}$ is the cross-correlation coefficient of large-scale structure tracer $I_i$ with the true lensing convergence, $\cl^{I_i I_i} $ is the power spectrum of $I_i$, and $\rho^{ij}$ are the cross correlation coefficients of tracers $I_i$ and $I_j$.

For the linear combination $I$, the correlation coefficient with CMB lensing is
\beq
\rho^2 = \sum_{ij} \rho_{i \kappa} (\rho^{-1})_{ij} \rho_{ j \kappa }.
\eeq
at each $l$.

For the linear combination of different CIB frequencies, we have all the information needed to calculate this correlation coefficient from the model for the CIB we use in this paper. We consider the combination of 545, 857, 217 and 353 GHz CIB maps, as well as the combination of only 545 and 857 GHz maps; the cross-correlation coefficient for the combination of four frequencies is shown in Fig.~1. For both linear combinations, we note that the improvement in correlation is quite small compared to individual CIB frequencies. This is expected: given how correlated the different CIB frequency channels we consider are (typically $\approx 95\%$, \cite{planckauto}), it is clear they must contain much of the same information.

Nevertheless, we can calculate the delensing improvement factor $\alpha$ for these linear combinations. As for individual frequencies, we obtain the improvement factors $\alpha$ by calculating the delensed residual B-mode power from Eq.~14. As can be seen in Table 1 and as expected from the correlation coefficient, the value of $\alpha$ for a noise-free experiment is 2.3 for both 545 GHz + 857 GHz and for all CIB frequencies combined (neglecting any CIB noise) -- not very different from the results of delensing with the $545$ GHz channel alone. 

Larger improvements may be possible if a wider range of less correlated frequencies is used, but we defer an investigation to future work.

\subsection{CIB Maps Combined with Lensing Reconstruction}

We now consider combining maps of the CIB with maps of lensing that have been reconstructed from CMB data. Such a combination could provide a better proxy for the true CMB lensing field and thus increase the delensing performance. To derive the delensing performance of such a combination, let us assume that our lensing reconstruction, for a certain instrumental noise level and beam, results in a lensing map with reconstruction noise $N_l^{\kappa \kappa}$, so that its power spectrum is
\beq
\cl^{\kappa_\mathrm{rec} \kappa_\mathrm{rec}} = \cl^{\kappa \kappa } +N^{\kappa \kappa}_l 
\eeq
We combine this reconstructed lensing map with a CIB map at one frequency (e.g., at 545 GHz), which we take to have a correlation coefficient $\rho^c_l$ with the \emph{true} lensing field. How efficient is this combination at delensing? To calculate this, we treat our lensing reconstruction as just another large-scale structure tracer $I_i$ and repeat our calculation of how well a combination of tracers can delens. The correlation matrices of Eq.~21 can be simply calculated to give the total correlation coefficient (at each $l$):
\beq
\rho^2_l =1 - \frac{(1-(\rho^c_l)^2) \frac{N^{\kappa \kappa}_l}{\cl^{\kappa \kappa}} }{(1-(\rho^c_l)^2)+ \frac{N^{\kappa \kappa}_l}{\cl^{\kappa \kappa}} }\nonumber
\eeq
\comment{as does the normalization which must be used for each multipole:
\beq
c_i = \frac{\{ 1-\rho_c^2, \rho_c \frac{N_l}{\cl^{\kappa \kappa}}  \sqrt{\frac{\cl^{\kappa \kappa}}{\cl^{I_c I_c}  }}\} }{  1-\rho_c^2+ \frac{N_l}{\cl^{\kappa \kappa}} }
\eeq}
Given our knowledge of the CIB-lensing correlation coefficient and our knowledge of the lensing reconstruction noise, we can calculate the improvement factor $\alpha$. 

We will first consider the case where the CMB lensing is estimated internally from the same survey that is being delensed. This requires an experiment with a small beam, such as our \emph{highRes} reference experiment. In Fig.~3, we include a curve showing the improvement factor $\alpha$ for delensing \emph{highRes} with the combination of 545 GHz CIB maps and quadratic estimator lensing reconstruction maps; moderate increases in $\alpha$ can be seen when delensing with this combination.

We now consider the delensing synergy of two different overlapping surveys -- an experiment similar to \lores ~(with low noise and low resolution that constrains large-scale B-modes and $r$, but cannot reconstruct lensing well) and an experiment similar to \hires ~(with high resolution that reconstructs $\kappa$ as a delensing tracer). This is a useful scenario because the high resolution experiment can be used to delens the low-resolution experiment if the surveys overlap \cite{namikawa}. The polarization noise level of the high resolution experiment $\Delta_P^{\mathrm{highRes}}$ will of course generally differ from that of the low resolution experiment $\Delta_P^{\mathrm{lowRes}}$. We show the delensing performance for such a scenario as a function of $\Delta_P^{\mathrm{lowRes}}$ in Fig.~4 (for \emph{highRes} noise values of $\Delta_P^{\mathrm{highRes}}= 2.8 \mu$K-arcmin and $\Delta_P^{\mathrm{highRes}}=7.1\mu$K-arcmin). The lensing reconstruction from \emph{highRes} is either used alone or combined with the 545 GHz CIB maps to delens \emph{lowRes}.

\begin{figure}[h!]
\subfigure{\includegraphics[width=\columnwidth]{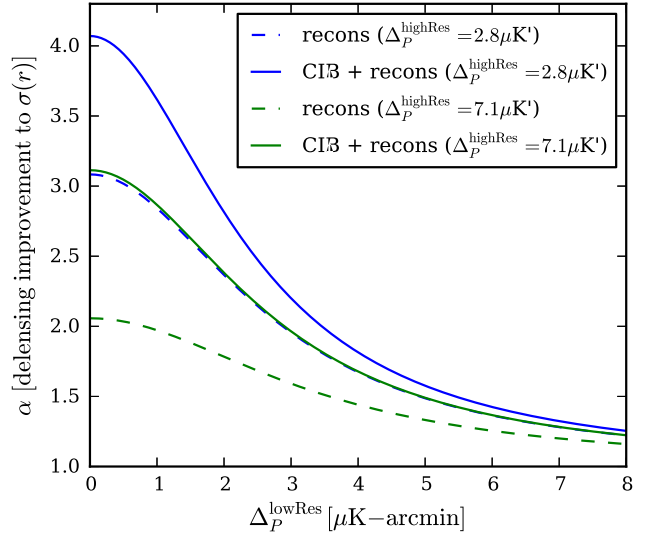}}
\caption{Delensing improvement factor $\alpha$ for delensing a low-noise/low-resolution CMB polarization experiment (e.g., Keck Array/BICEP3 -- for us, the \emph{lowRes} reference experiment) with lensing maps reconstructed from an overlapping higher resolution CMB experiment (e.g., ACTPol/SPTpol/POLARBEARI/II -- for us, the \emph{highRes} reference experiment). The lensing reconstruction maps are used either alone or in combination with 545 GHz CIB maps. We assume all constraints on $r$ are from \lores, which has a noise level given by $\Delta_P^{\mathrm{lowRes}}$. The external CMB lensing reconstruction used for delensing is assumed to be reconstructed with a quadratic estimator from \hires, which has a noise level $\Delta_P^{\mathrm{highRes}}$. Shown are values of $\Delta_P^{\mathrm{highRes}} = 2.8 \mu $K-arcmin (upper dashed line) or $\Delta_P^{\mathrm{highRes}}= 7.1 \mu $K-arcmin (lower dashed line). The solid lines correspond to delensing with optimal combinations of lensing reconstruction and the 545 GHz Planck CIB map (with the same ordering by \emph{highRes} noise). It can be seen that adding CIB data to a lensing reconstruction map greatly increases the delensing performance, and that delensing a low-resolution, low-noise experiment with a combination of CIB maps and overlapping intermediate-noise external lensing reconstruction is promising. We assume that the E-mode map used for delensing is signal dominated, which is likely to be the case if we can combine high- and low-resolution data.\comment{TODO: maybe mention lmax BB and lmax EE used to generate the plot?}}
\end{figure}

We find that the combination of CIB and reconstructed lensing from high-resolution CMB maps, even with intermediate noise levels of $\Delta_P^{\mathrm{highRes}} \approx 3\mu$K-arcmin, is quite effective as a delensing tracer, and is able to improve the detection errors by close to a factor $4$ for a lensing-limited survey. The addition of the CIB results in a significant improvement in delensing performance compared to reconstruction with \hires ~alone; noisy $\Delta_P^{\mathrm{highRes}} = 7.1 \mu $K-arcmin reconstruction coadded with CIB data delenses as well as $\Delta_P^{\mathrm{highRes}} = 2.8 \mu $K-arcmin reconstruction without the CIB. This increased performance is due to the fact that lensing reconstruction at intermediate noise levels does not image small lenses well; adding the CIB provides information on these small lenses and increases the small-scale correlation of the coadded delensing tracer with the true lensing convergence. The combination of a low noise B-mode search (similar to \lores) that cannot reconstruct lensing well with a higher resolution experiment (such as \hires), even at a higher noise level, is thus quite promising for delensing, especially when this external lensing reconstruction is combined with CIB data. (We note that our calculations are only relevant when the constraint on $r$ from \lores~is more powerful than the constraint from \hires ~alone, e.g., because the latter has higher noise or has not been optimized to measure large-scale polarization).

Additional improvements might be obtained by combining CIB data with other tracers of large-scale structure, such as galaxy surveys (or other CIB maps over a wider frequency range), in order to improve the overlap of the combination's redshift origin with the lensing kernel. However, we leave such an investigation to future work.

\section{Potential Limitations of Observational CIB Data as a Delensing Tracer}

\subsection{Uncertainty in the CIB Cross- and Auto-Power Spectra, and Prospects for Self-Calibration}

In our calculations of section 3, we assumed a model (albeit one based on measurements) for the lensing cross- and auto-spectra $\cl^{\kappa I}$ and $\cl^{II}$ of the CIB. However, though good measurements are available, the true cross-and auto-spectra of the CIB fields are of course not exactly known. This is potentially important for estimating delensing performance, not just because forecasts may be slightly inaccurate, but because the true CIB lensing-cross- and auto-spectra determine the residual B-mode power spectrum (the power spectrum of the lensing B-mode that remains after delensing) as in Eq.~12. Because the CIB spectra affect the residual B-mode power spectrum, insufficiently constrained CIB spectra can degrade constraints on $r$. For example, if the CIB cross-correlation is in fact smaller than in the fiducial model, the reduced correlation leads to an ``excess'' of residual lensing B-mode power compared to the fiducial model, which could be confused with the presence of primordial tensor B-modes. We illustrate this by constructing 10 different CIB cross-correlation curves, all of which are consistent with current Planck measurements, and calculating the residual B-mode power spectrum from Eq.~12 for each such model. (To construct each curve, we perturb every CIB-lensing-cross model bandpower by a random amount drawn from a Gaussian of width given roughly by the size of the measurement errors from \cite{planckciblensing}; see later for more discussion of our error assumptions). In Fig.~5 we show the difference between the residual B-mode spectrum for each of these 10 alternative ``models'' and the residual B-mode spectrum for the fiducial model. We also show a primordial tensor B-mode power spectrum corresponding to $r=0.001$ for comparison. Though the differences in the residual B-mode power are small (of order 5\% of the total power), it can be seen that they are comparable in size to primordial B-modes when $r\sim 0.001$; uncertainty in the CIB spectra could thus limit the power of CIB delensing for B-mode surveys that are capable of such high precision constraints on $r$ (e.g., are sufficiently large).
\begin{figure}[h!]
\includegraphics[width=\columnwidth]{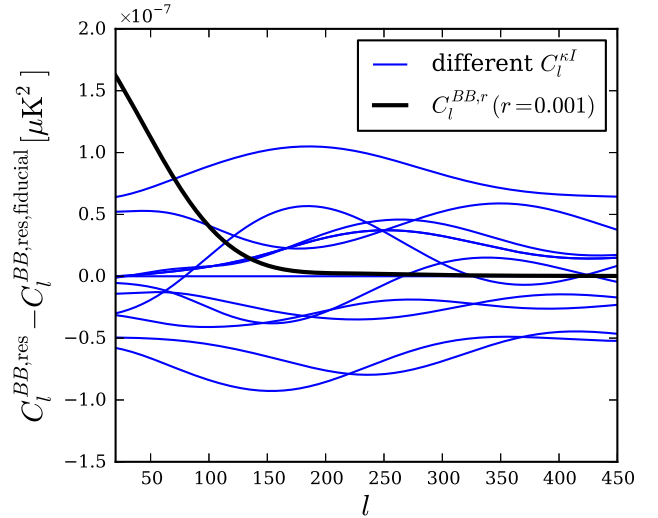}
\caption{Illustration of the possible impact of uncertainty in the true CIB-lensing cross-correlation on the residual (delensed) B-mode power spectrum. We construct 10 different possible ``models'' of the true cross-power by perturbing bandpowers of our fiducial model by a random amount drawn from a Gaussian, Planck-like error distribution. Each model cross-power is consistent with current measurements but results in a different residual B-mode power spectrum; the excess residual B-mode power compared with our fiducial model is shown in blue. For comparison, the power spectrum of primordial B-modes with an amplitude corresponding to $r=0.001$ is shown. It can be seen that variation in the uncertain CIB spectra can be degenerate with the presence of primordial B-modes, especially if only the largest scales of the B-mode polarization are measured.}
\end{figure}

We now seek to quantify the effects of uncertainty in the CIB spectra on the delensing procedure more rigorously. We begin by writing a theoretical model for the full delensed B-mode power spectrum, $\cl^{BB,\mathrm{del}}$, as:
\beq
 \cl^{BB,\mathrm{del}}(r,a_i,b_i) = \cl^{BB,r}(r) +  {C}_l^{BB,\mathrm{res}}[\cl^{\kappa I}(a_i), \cl^{II}(b_i)]\nonumber
\eeq
where we have denoted the residual lensing B-mode power as a function of the CIB spectra, and we have assumed a measurement without noise bias. We have also assumed that, in turn, the parameters $a_i, b_i$ (which we will later define as effectively parametrizing the amplitude of CIB bandpowers) completely describe the true CIB cross- and auto-spectra. As illustrated previously, uncertain CIB spectral parameters $a_i, b_i$ could potentially be degenerate with $r$ and thus reduce the performance of CIB delensing. To quantify this limitation, we will again forecast constraints on $r$, but now marginalize them over the uncertain true CIB spectra (i.e., the parameters $a_i, b_i$).

First, we discuss how the CIB spectra can be parametrized for delensing analyses. While considerable progress has been made in modeling these spectra, for cosmological constraints on $r$ we believe it is more robust if cross- and auto-spectra can be completely constrained from measurements without assuming a certain model for the CIB properties and spectra. This is especially important as foreground residuals may also contribute to the spectra and may not be simple to model. We thus propose to constrain $r$ and the CIB spectral parameters $a_i, b_i$ simultaneously from measurements of $C_l^{BB,\mathrm{del}}$ as well as measurements of the CIB spectra ${C}_l^{\kappa I}$ and ${C}_l^{II}$, and then marginalize constraints on $r$ over $a_i, b_i$. ${C}_l^{II}$ should ideally be measured with the exact CIB map used for delensing, including any foreground (e.g., galactic dust) residuals present in the map. As long as fiducial model spectra are close to the data, they should be used for the filtering function $f$ of Eq.~12. This preserves linearity and is only slightly suboptimal, because any correction to the delensing performance is second order in an error in $f$. For such an empirical approach, we must have a very general parametrization of the CIB spectra. As indicated previously, we will here assume that the CIB lensing-cross- and auto powers can be completely described by rescaling different $l$-ranges of the fiducial CIB spectra $ \tilde{C}^{\kappa I},  \tilde{C}^{I I}$, similar to rescaling effective bandpowers:
\beqn
&& C^{\kappa I}_{{l}} (a_{i}) \equiv  \tilde{C}^{\kappa I}_{{l}}\left(1 + a _{i} U_i({l}) \right) , \nonumber \\&& C^{I I}_{{l}} (b_i) \equiv  \tilde{C}^{I I}_{{l}}\left(1 + b _{i} U_i({l}) \right) \eeqn
where a sum over $i$ is implicit, and $U_i$ are top-hat functions that are unity within an $l$-range corresponding to the $i$th bin (we describe our binning in later sections), and zero otherwise, i.e.:
\beq
\delimiterfactor=1200 
U_i(l)  = \left\{%
\begin{array}{ll}
1 &\textrm{if }l_\mathrm{i,min} <l<l_\mathrm{i,max},\\
0 &\textrm{otherwise}.
\end{array}%
\right.
\eeq
(though there are other possible, equally complete, parametrizations). In writing this model, we assume that the true CIB spectra in our data are sufficiently smooth that they can be described without significant losses by this discrete parametrization when integrated over the kernel of Eq.~5 (and Fig.~1), though this assumption appears reasonable unless the binning is very coarse.

Assuming that measurements of both CIB spectra and the B-mode power spectrum are available, and that we are fitting both $r$ and the CIB spectral parameters simultaneously, we now investigate to what extent uncertain ${C}_l^{\kappa I}$ can degrade the constraints on $r$. We will neglect uncertainty in the CIB auto-power, because we assume that the raw auto-spectrum of the CIB map (including any residual foregrounds) can be measured with Planck data much more accurately than the lensing cross-spectrum. This is currently the case for the relevant high frequency CIB maps at 545 and 857 GHz (despite the smaller area on which the auto-spectrum is measured), and is typically a good approximation because the lensing reconstruction maps are expected to have significantly higher noise than the CIB maps. (Furthermore, the residual B-mode is twice as sensitive to changes in the cross-power than to the auto-power, as can be inferred from Eq.~12).  To investigate how much constraints on $r$ are degraded by uncertain ${C}_l^{\kappa I}$, we perform a Fisher forecast for the parameters $\theta_p = (r, a_i)$, obtaining a constraint on $r$ marginalized over the CIB-lensing cross-power parameters $a_i$. We construct the Fisher matrix as
\beqn
F_{pq} &=& \sum_{l_{a} =l^{BB}_{\mathrm{min}}}^{l^{BB}_{\mathrm{max}}}  \sum_{l_{b} =l^{BB}_{\mathrm{min}}}^{l^{BB}_{\mathrm{max}}}   \frac{\partial {C}_{l_a}^{BB,\mathrm{del}}}{\partial \theta_p } \left[ \mathrm{Cov}^{BB,BB}\right]_{l_a, l_b} ^{-1} \frac{\partial {C}_{l_b}^{BB,\mathrm{del}}}{\partial \theta_q }\nonumber \\&+&  \sum_j \frac{\frac{\partial {C}_j^{\kappa I}}{\partial \theta_p } \frac{\partial {C}_j^{\kappa I}}{\partial \theta_q }}{(\Delta C_j^{\kappa I})^2}
\eeqn
where the cross-power has been binned into bandpowers $j$ which are identical to the effective bins of the $a_i$ parametrization. To construct this Fisher matrix, we must evaluate derivatives of $C_l^{BB,\mathrm{del}}$ and $C_l^{\kappa I}$ with respect to all parameters $\theta_p$, and estimate the measurement errors (covariance matrices) of the measurements of ${C}_{l}^{BB,\mathrm{del}}$ and ${C}_{l}^{\kappa I}$. 

We begin by calculating how changing the assumed CIB cross-spectrum via $a_i$ affects the delensed residual B-mode power. Recalling our expression for the residual B-mode power (Eq.~12), we  can insert our parametrization of the CIB spectra (Eq.~25) in terms of $a_i, b_i$ into this expression (keeping the weight $f$ fixed to the fiducial model) to obtain:

\beqn
&&{C}_{l}^{BB,\mathrm{res}}=   \int \frac{d^2 \bl'}{(2 \pi)^2}  W^2 (\bl,\bl') 
 C^{EE}_{l'} C^{\kappa \kappa}_{|\bl-\bl'|} \times \\
&&  \left[1 - \frac{C^{EE}_{{l'}} ~\tilde{\rho}^2_\mathrm{|\bl-\bl'|} }{C^{EE,N}_{{l'}}}  \left(1 + 2 a_i U_i(\mathrm{|\bl-\bl'|}) - b_i U_i(\mathrm{|\bl-\bl'|}) \right)\right] \nonumber
\eeqn
where $\tilde \rho$ is the correlation coefficient for the fiducial model. We can now calculate the derivatives of the delensed B-mode power spectrum (Eq.~28) with respect to the $a_i$ parameters. Using the chain rule it can be seen that
\beq
\frac{\partial {C}_{l_{a}}^{BB,\mathrm{del}}}{\partial a_i} = -2  \sum_{{l \in l_i}} \frac{\partial {C}_{l_{a}}^{BB,\mathrm{lens, N}}}{\partial C_{l}^{\kappa \kappa}} C_{l}^{\kappa \kappa}  \tilde{\rho}^2_{l} 
\eeq
where we have rewritten the expression in terms of the full-sky derivative $ {\partial {C}_{l_{a}}^{BB,\mathrm{lens}}}/{\partial C_{l}^{\kappa \kappa}} $ with discrete multipoles $l$. We evaluate the lensing derivative in this equation by differentiating the analytical full-sky expression of \cite{huCMB}  for $C^{BB,\mathrm{lens}}_{l_{a}}$, which is correct to leading order in $C_{l}^{\kappa \kappa}$. The $N$ superscript on the derivative indicates that when the beam is large and the E-mode is not signal dominated, the derivative must be evaluated after applying a filter $C_l^{EE}/C_l^{EE,N}$ to the E-mode power. The derivatives of the CIB-lensing cross-power with respect to $a_i$ and the B-mode power with respect to $r$ are much simpler
\beq
\frac{\partial {C}_j^{\kappa I}}{\partial a_i} = \delta_{ij} \tilde C_j^{\kappa I};~~
\frac{\partial {C}_l^{BB,\mathrm{del}}}{\partial r} = C_l^{BB, r}(r=1)
\eeq
where we have noted that $C_l^{BB, r} = r \times C_l^{BB, r}(r=1)$ for all $r$.

We now turn to determining errors on the measurements. For a measurement of the B-mode power spectrum, the errors are determined by the covariance matrix:
\beqn
\mathrm{Cov}^{BB,BB}_{l_a, l_b}  &=& \frac{2 \delta_{l_a, l_b}}{\fsky (2 l_a +1 )} (C_{l_a}^{BB,\mathrm{res}}+N_{l_a}^{BB,\mathrm{lens}})^2  \\ &+&  \frac{1}{\alpha_0^2}\sum_{l}  \frac{\partial {C}_{l_a}^{BB,\mathrm{lens}}}{\partial {C}_{l}^{\tilde E \tilde E}} \left[ \mathrm{Cov}^{\tilde E \tilde E, \tilde E \tilde E}_{l, l} \right] \frac{\partial {C}_{l_b}^{BB,\mathrm{lens}}}{\partial  {C}_{l}^{\tilde E \tilde E} }\nonumber \\ &+&  \frac{1}{\alpha_0^2} \sum_{l}  \frac{\partial {C}_{l_a}^{BB,\mathrm{lens}}}{\partial {C}_{l}^{\kappa \kappa}} \left[ \mathrm{Cov}^{\kappa \kappa, \kappa \kappa}_{l, l} \right] \frac{\partial {C}_{l_b}^{BB,\mathrm{lens}}}{\partial  {C}_{l}^{\kappa \kappa} }\nonumber
\eeqn
where the second and third lines give the dominant part of the non-Gaussian contribution to the covariance \citep{benoit}. When calculating delensed constraints, we have assumed that the off-diagonal parts of the covariance matrix are reduced by the constant delensing improvement factor $\alpha_0 = \langle C_l^{BB,\mathrm{lens}} \rangle_{l<100}/ \langle C_l^{BB,\mathrm{res}}\rangle_{l<100}$, neglecting any scale-dependence. We also neglect the weak dependence of  $C_{l}^{BB,\mathrm{res}}$ and $\alpha_0$ on $a_i$ given current constraints. We assume that the lensing covariance itself is Gaussian
\beq
\mathrm{Cov}^{\kappa \kappa,\kappa \kappa}_{l_a, l_b} = \frac{2}{\fsky (2 l_a +1 )} (C_{l_a}^{\kappa \kappa})^2 \delta_{l_a, l_b}
\eeq
and that an analogous expression holds for $\tilde E$, the unlensed E-mode. We again use first order analytical derivatives of the B-mode power spectrum in constructing the off-diagonal part of the B-mode covariance matrix (Eq.~31), unlike in \cite{benoit}, where full numerical derivatives are evaluated. Higher order corrections can cause changes to the off-diagonal covariance, but as we find that the omission of the non-Gaussian covariance contributions only causes negligible (much less than $10\%$) changes to our final results, our simple approximation of the off-diagonal covariance appears sufficiently accurate for our estimates.

We must also make an estimate of the measurement uncertainty on the CIB-lensing cross-spectrum over the range of scales that contributes most to the signal-to-noise. We consider two cases for the errors on the CIB-lensing cross-power, corresponding to current or near-future measurements: first, a constraint similar to the currently available Planck CIB-lensing cross-correlation data, second, a substantial improvement on this with the errors halved. We will make a simple model of the Planck measurement errors as follows. We assume 23 bandpowers, of width $\Delta l = 63.5$ (half the width of Planck bandpowers), with the first bandpower centered at $l = 72.5$, i.e. roughly spanning $60<l<1500$ (we maintain our cutoff below $l = 60$). We take the $a_i$ parameters to have the same effective bins, but will revisit the choice of effective bandpower width at the end of this section. The magnitude of the errors on each of these bandpowers is assumed to be a constant $\Delta(2 l_j C_j^{\kappa I}/\sqrt{2}) = 0.02$uK.sr, which is a good approximation to the the errors on the Planck cross-correlation (\cite{planckciblensing}, e.g., Table 2 and Figure 5; optimistically, we only use the statistical part of the errors). This corresponds to a $\approx 50\sigma$ measurement of the cross-power, with fractional errors on the bins ranging from $6-18\%$. The other, more futuristic, scenario we consider corresponds to the same assumed measurements but with the errors halved. We assume that any signal covariance between the delensed B-mode measurement and the measurement of the lensing-CIB cross-correlation is negligible, because the Planck lensing measurements are still reconstruction-noise-dominated on the scales most important for delensing (and in addition, the areas of sky used for the two measurements will often differ).

With our knowledge of the errors on both observables (B-mode power and the CIB-lensing cross power) and the dependence of the observables on the parameters $\theta_p = (r, a_i)$, we construct the Fisher matrix of Eq.~27 and obtain the marginalized error on $r$ as follows:
\beq
\sigma_{\mathrm{marginalized/delensed}}(r)=\sqrt{(F^{-1})_{r r}}
\eeq

We can now compare two cases. First, a case in which we do not delens; in this case, there is no reduction of the lensing noise in a measurement of the B-mode power, but there is also no delensing-induced degeneracy between $r$ and $a_i$. Second, a case in which we have delensed; this reduces the B-mode noise by up to $\approx 2.2$, but also introduces a degeneracy of $r$ with the set of uncertain CIB parameters $a_i$. To quantify this tradeoff, we define the marginalized improvement factor as the ratio of the errors on $r$ for these two cases, now including the effects of marginalizing over the uncertain correlation parameters $a_i$ in the delensed error:
\beq
\alpha_{\mathrm{marginalized}} = \sigma_0(r) / \sigma_{\mathrm{marginalized/delensed}}(r)
\eeq
We can plot this ratio as a function of $\fsky$, initially assuming that due to foregrounds and systematics, the largest scales below $l^{BB}_\mathrm{min}=20$ (and hence any tensor B-modes from reionization) cannot be measured. 

To gain intuition, we begin by neglecting instrumental noise in the B-mode measurement, and simply impose a sharp cutoff $l^{BB}_\mathrm{max}$ beyond which B-modes are not measured (though we later investigate more realistic scenarios with different beam and noise specifications). The results are shown in Fig.~6, for different ranges of $l$ which contribute to the B-mode power spectrum measurement ($l^{BB}<l^{BB}_\mathrm{max} = 100$, $l^{BB}_\mathrm{max} = 300$, $l^{BB}_\mathrm{max} = 450$) as well as two different assumptions for the errors on the measurement of the CIB-lensing correlation (Planck, Planck/2).

\begin{figure}[h!]
\includegraphics[width=\columnwidth]{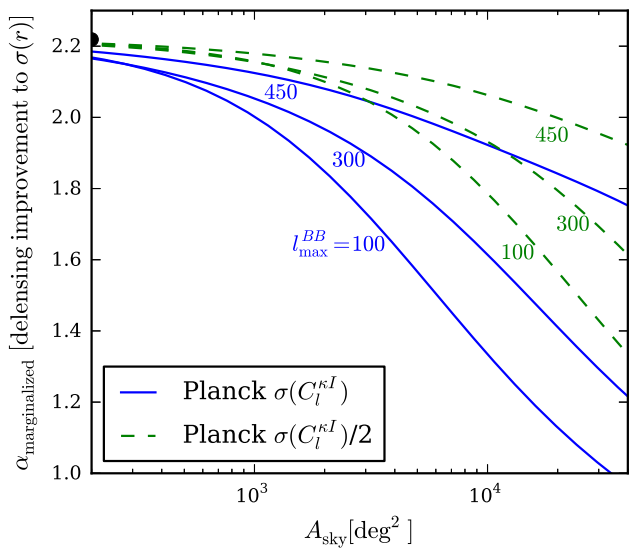}
\caption{Delensing improvement factor $\alpha$ on a constraint on $r$, as a function of survey area -- now marginalized over the uncertain CIB-lensing cross-spectrum. If the CIB-lensing correlation were known perfectly, $\alpha$ would be constant with area, as both delensed and non-delensed errors would change by the same factor as the area is varied. However, marginalization changes this for large surveys, as can be seen from the falling blue solid lines (corresponding to Planck-like errors on the CIB-lensing correlation) and green dashed lines (corresponding to Planck errors halved). The degradation in delensing performance seen is due to a delensing-induced degeneracy of $r$ with the CIB-lensing cross-spectrum; for large, powerful surveys, this degeneracy significantly increases the delensed errors on $r$ upon marginalization. To gain insight, we have here neglected instrumental noise, and have simply added a sharp cutoff $l^{BB}_\mathrm{max}$ to which the B-mode power is measured. For each set of lines, the lower line corresponds to $l^{BB}_\mathrm{max} = 100$, the middle line to $l^{BB}_\mathrm{max} = 300$ and the upper line to $l^{BB}_\mathrm{max} = 450$. The constant value of $\alpha$ when there is no uncertainty in the CIB spectra is also shown as a black semicircular point at the top left of the figure. As the resolution of the B-mode power spectrum measurement is increased from $l^{BB}_\mathrm{max} = 100$ to $l^{BB}_\mathrm{max} = 450$, it can be seen that the degeneracy of $r$ with the cross-correlation parameters is broken as the CIB-lensing correlation is increasingly self-calibrated by the B-mode power spectrum at small scales.}
\end{figure}

\begin{figure}[h!]
\includegraphics[width=\columnwidth]{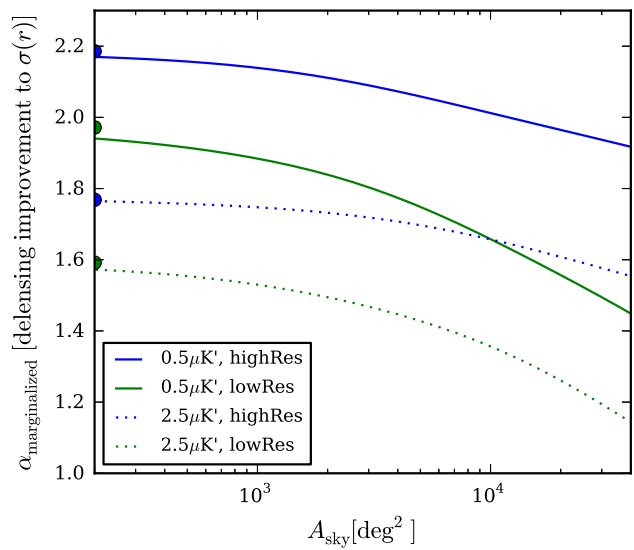}
\caption{Delensing improvement factor $\alpha$ on a constraint on $r$ as a function of survey area, marginalized over the CIB-lensing cross-correlation, as in Fig.~6. Instead of the sharp $l^{BB}_\mathrm{max}$ cutoffs used to gain understanding in Fig.~6, we show more realistic results for different noise and beam specifications (\emph{highRes} corresponds to a $1.4'$ beam, \emph{lowRes} a $30'$ beam). Planck-like errors on the CIB-lensing correlation are assumed. For reference, the ideal improvement factors when there is no uncertainty on the CIB spectra are indicated with solid circles at the left of the plot. Our results again confirm that marginalizing over the uncertain CIB spectra does not greatly degrade constraints for small surveys or for surveys which can resolve small B-mode scales (and can hence self-calibrate). For many surveys, CIB delensing thus does not require the assumption of a model for the CIB spectra. }
\end{figure}

We first discuss our calculation when we restrict ourselves to $l^{BB}<100$. It can be seen that as the area of the survey becomes large, $\alpha_{\mathrm{marginalized}} = \sigma_0(r) / \sigma_{\mathrm{marginalized/delensed}}(r)$ falls rapidly. As the area is increased, the marginalized $\sigma_{\mathrm{marginalized/delensed}}(r)$ must therefore become much greater than its value for fixed CIB parameters. (Note that $\sigma_0$, the error when not delensing, is unaffected by marginalization). We can understand this as follows. With larger survey area, all constraints on $r$ become tighter; with stronger constraints, the sensitivity of a constraint on $r$ to cross-power uncertainty increases, because smaller $r$ can more easily be mimicked by uncertain cross-powers as in Fig.~5. As large, powerful surveys are more sensitive to cross-power uncertainty, marginalization over the uncertain cross-powers leads to a significant increase of errors on $r$ when surveys are large. For a survey larger than $\sim 30000$ square degrees, this sensitivity is such that it is no longer worth delensing if this comes at the cost of introducing a degeneracy of $r$ with uncertain CIB cross-powers. (However, note that our current delensing weighting is not optimized to account for this marginalization.)

The conclusion that it is not worth CIB delensing the largest surveys changes either if the cross-correlation measurements are greatly increased in accuracy, or if we use the constraining power of smaller scales in $C_l^{BB,\mathrm{del}}$. Increasing $l_\mathrm{max}$ to 450, we can see that now the improvement upon delensing remains close to 2.2, the value we obtained when neglecting uncertainty in the CIB-lensing cross-power, even for large surveys. An explanation for this observation is that the degeneracy of the cross-correlation parameters and $r$ is being broken by measurements of the small-scale B-modes; effectively, the higher-$l$ B-mode power spectrum measurements are determining the CIB-lensing correlation and are fixing the amplitude of the low-$l$ lensing B-mode residual, uncertainty in which would otherwise be degenerate with $r$. We confirm this by noting that the constraints on any $a_i$, marginalized over $r$ and holding the other $a$ parameters at their fiducial values, are tightened significantly as $l_\mathrm{max}$ is increased beyond $l_\mathrm{max}=100$ (whereas for $l_\mathrm{max}=100$, including the B-mode measurements does not tighten the constraints on $a_i$). It is possible that this self-calibration could also have been equivalently achieved via noisy lensing reconstruction using only $l = 100-450$ information (though a reconstruction may have insufficient signal-to-noise per mode to delens, it may still allow a reasonable measurement of the cross-correlation given enough sky area).

As expected, the constraints on $r$ are much less affected by marginalizing over uncertain CIB-lensing cross-powers if the cross-powers are measured twice as accurately by a futuristic lensing survey.

To make our analysis more realistic, we now repeat our calculations for the \emph{highRes} and \emph{lowRes} reference experiments introduced earlier, now including instrumental noise in our calculation. We show results for both experiments, at two different noise levels, in Fig.~7. We again see that for small surveys, marginalizing over the uncertain CIB spectra barely degrades constraints from their ideal values. It can also be seen that the degradation with sky area is less severe for high-resolution experiments  (blue curves) than for low-resolution ones (green curves). This is consistent with the $l^{BB}_\mathrm{max}$-dependence discussed in the context of Fig.~6, and is due to the fact that high-resolution experiments can use the high-$l$ B-modes for better self-calibration. For \lores, the improvement factor falls, though generally does not degrade as much with sky area as when a sharp cutoff is applied (as in Fig.~6), because even high-$l$ B-modes that are individually noisy contribute when averaging over many modes. Our results confirm that, for a range of realistic surveys (in particular, small surveys and high resolution surveys), a specific CIB spectrum need not be assumed; the spectra can instead be marginalized over without a significant loss of delensing performance.

We will briefly discuss how our results change if we now assume that the reionization B-modes at $l^{BB}<20$ can be measured. In this case, we do not find a substantial reduction of delensing performance when marginalizing over the CIB spectra, even for very large surveys. We explain this as follows: when reionization B-modes can be measured, most of the constraining power on $r$ arises from the largest scales $l^{BB}<20$; on these scales, the effects of uncertain CIB spectra are much less important (as they are subdominant to the large lensing B-mode variance).

\comment{For small surveys and surveys that can constrain reionization B-modes, the uncertainty in the CIB spectra is negligible in general; if the $l = 100-450$ (or to some extent $l = 100-300$) B-modes can be measured, uncertainty in the CIB spectra will not even substantially degrade the delensing performance for large surveys. Any degradation is further reduced if improved measurements of the CIB-lensing cross-power become available. We conclude that, for many surveys, CIB delensing can be performed effectively without assuming a fixed model for the CIB properties and spectra.}

Finally, we briefly comment on our use of effective bandpowers $a_i$ as a very simple non-parametric description of the CIB cross-power for forecasting and marginalization purposes (see \cite{joachimi} for a similar analysis in a different context). We focus first on choice of their width. It should be noted that to obtain the residual B-mode power of Eq. 14, the correlation is integrated over the kernel plotted in Fig.~1. Our bandpower widths are chosen to be somewhat smaller than the scale on which this kernel varies strongly. Still narrower bins are thus unnecessary to characterize effects on the residual B-mode power, as any sub-bandpower features in the CIB cross-correlation will be averaged out by the integration. Of course, since most physical processes should lead to smooth features in the cross-power, the use of wider bins, different general parametrizations that encode smoothness (such as a limited set of Fourier or polynomial coefficients) or Gaussian processes may be well-motivated. However, investigating the impact of our choice of bandpower width by doubling (and halving) their width, we find that our results are effectively unchanged. We thus believe that marginalizing over our current set of CIB parameters should give a conservative, though reasonable, estimate of the degradation of constraints on $r$.
\subsection{Foregrounds and Non-Gaussianity of the CIB Maps}
We have argued in section 3.1 that our forecasts are realistic for many surveys, despite the presence of galactic dust foregrounds in the CIB maps. We have also argued that the raw CIB auto-spectra (including residual foreground power) can be measured accurately, so that additional foreground power should not bias or degrade results if $r$ and the true CIB power spectra are simultaneously constrained. However, there are two potential limitations of our analysis, which will need to be revisited in future work.

First, we have assumed the CIB maps to be Gaussian, whereas some non-Gaussianity is in fact expected, both due to intrinsic non-Gaussianity in the CIB field and due to residual foregrounds or secondary anisotropies. 
If we assume perfect CMB cleaning, any non-Gaussianity in $I$ is not expected to lead to biases, as the mean residual B-mode power spectrum is obtained from two powers of the CIB field $I$. Though non-Gaussianity in $I$ could potentially affect the size of the error bars, integration of many independent modes is expected to reduce the already-moderate non-Gaussianity present in the two-halo-term dominated $l=200-500$ CIB scales most important for delensing. However, this will need to be quantified in future work.

Second, we have assumed that the E and B-mode maps have been cleaned successfully using multifrequency data, so that any foreground residuals can be neglected. If the CMB component separation and cleaning is imperfect, additional complications can arise. Though CIB anisotropies are expected to be negligible in the large-scale CMB polarization (as galaxy polarization directions add incoherently), any remaining galactic dust intensity features in the CIB may be correlated with polarized galactic dust residuals in the CMB E and B maps. In principle, this could lead to additional Gaussian and non-Gaussian contributions to the delensed B-mode power spectrum and its error. While dust residuals are expected to be small in the cleaned CMB and CIB maps, and the galaxy-dominated low-$l$ modes (which overlap with the scales used to constrain $r$) can be excluded from the delensing CIB map without significant losses, the magnitude of any biases and corrections are currently uncertain.

Therefore, though simple arguments suggest that the impact of foreground residuals and non-Gaussianity on the delensing procedure is likely to be moderate, a more detailed investigation of these effects, with simulations and data, will be required in future work. We note that further investigation of some of these potential issues is also well-motivated for internal delensing methods.

\section{Conclusions and Outlook}
We have examined in detail the methods and performance of CIB delensing, including combinations of different frequencies as well as combinations of CIB maps with lensing reconstruction. Our results show that CIB delensing can remove more than $50\%$ of the lensing power and can hence lower the noise of a B-mode measurement by a factor of $\approx 2.2$, and significantly more when CIB maps are combined with lensing reconstruction. Marginalizing over the true CIB spectra, we find that uncertainty in the CIB spectra does not degrade delensing performance significantly for small surveys, and only moderately reduces delensing performance for large surveys with high angular resolution. This confirms that, in most cases, CIB delensing can be performed without assuming a model for the CIB spectra. However, further investigation into systematic errors from foregrounds and non-Gaussianity will be required in future work.

At noise levels of $\approx 1\mu$K-arcmin (as targeted by CMB stage-IV) or below, internal delensing by reconstructing the lensing field from the CMB itself is significantly more powerful than CIB delensing. However, for many experiments that are currently planned, CIB delensing appears competitive with internal delensing. Even experiments such as COrE, LiteBIRD or PRISM \citep{core,litebird,prism} are expected to have polarization noise levels of $2 \mu$K-arcmin or higher (especially after foreground cleaning). At these noise levels, internal delensing does not perform much better than CIB delensing (e.g., in the case of iterative delensing with $2 \mu$K-arcmin noise, the improvement factor is $\alpha \approx 2.7$ from \cite{smithdelensing}, whereas CIB delensing gives $\alpha \approx 1.9$). Perhaps most significantly, many current and upcoming surveys in the next decade do not have the high angular resolution required to delens well internally (e.g., Keck Array, BICEP3, LiteBIRD; \cite{kernasovskiy, bicep3, litebird}), so that external delensing is the only feasible method; for these experiments, CIB delensing is a promising option. 

CIB delensing can significantly strengthen constraints on tensor B-mode polarization while requiring only minimal assumptions about CIB properties. It is thus a robust and promising method for enhancing the upcoming generation of CMB polarization surveys.
\acknowledgements{We thank R.~Keisler, U.~Seljak, A.~van Engelen, O.~Dor\'{e}, J.~C.~Hill, E.~Linder, S.~Ferraro, J.~Errard and J.~Dunkley for helpful discussions and/or feedback on the paper draft. We are particularly grateful to G.~Lagache for making the CIB halo model from \cite{planckauto} available. BDS was supported by a Fellowship from the Miller Institute for Basic Research in Science at the University of California, Berkeley. BDS also acknowledges the hospitality of the Beecroft Institute of Particle Astrophysics and Cosmology at Oxford University, where part of this work was completed.}

\appendix
\section{Combining Tracers for Delensing}

We now derive the optimal combination of multiple fields which trace large-scale structure, such as CIB maps measured in different frequency channels (without assuming redshift distributions for these fields), in order to maximize delensing performance. This optimal combination can be described by writing the tracer $I$ as a linear combination of different fields
\beq
I(\hat{\mathbf{n}}) = c_i I_i 
\eeq
where $I_i$ are the different fields to be combined, $c_i$ are the linear combination coefficients and repeated indices are summed over.

To construct a field which is optimal for delensing and cross-correlating, we now solve for the coefficients $c_i$ which maximize the correlation coefficient:
\beq
\rho^2_l = \frac{ (\cl^{\kappa I})^2 }{\cl^{\kappa \kappa}  \cl^{II}} = \frac{  \left[ c_i \langle I_i \times \kappa \rangle \right]^2 }{\cl^{\kappa \kappa}   c_i c_j \langle I_i \times I_j \rangle} =  \frac{  \left[ c_i   \cl^{\kappa I_i}   \right]^2 }{\cl^{\kappa \kappa}   c_i c_j  \cl^{I_i I_j} }
\eeq
We maximize by taking the derivative of $\ln \rho^2$ with respect to $c_a$
\beqn
0=\frac{\partial}{\partial c_a} \ln(\rho^2) &=& \frac{\partial}{\partial c_a} 2 \ln  \left[ c_i \cl^{\kappa I_i} \right] - 
\frac{\partial}{\partial c_a} \ln \left[  c_i c_j \cl^{I_i I_j}  \right]\nonumber \\ &=&
\frac{2 \cl^{\kappa I_a}  }{   c_i \cl^{\kappa I_i}   } - \frac{ 2   c_i   \cl^{I_i I_a}  }{c_i  c_j \cl^{I_i I_j}   }
\eeqn
We note that this equation can be rewritten as a matrix equation involving the optimal vector of coefficients $c_a$ or $\mathbf{c}$:
\beq
0= 2\frac{ \mathbf{k}}{\mathbf{c} \cdot  \mathbf{k}} - 2\frac{ \mathbf{Q}~\mathbf{c} }{\mathbf{c} \cdot  [\mathbf{Q}~\mathbf{c}]},
\eeq
where we have defined the vector $\mathbf{k}$ and the matrix $\mathbf{Q}$ as follows:
\beq
k_i=\cl^{\kappa I_i}  ,
\eeq
\beq
Q_{ij} = \cl^{I_i I_j}  
\eeq
It can thus be seen that the correlation coefficient is maximized when
\beq
\mathbf{k} = \mathbf{Q}~\mathbf{c} \times n
\eeq
where $n$ is an arbitrary normalization factor, set by Eq.~13. For clarity, we will assume $n=1$ for now, and will verify that our normalization is correct by checking at a later stage that the normalization Eq.~13 is satisfied by this choice.

The optimal coefficients are thus given by
\beq
\mathbf{c} = \mathbf{Q}^{-1}~\mathbf{k}
\eeq
Rewriting the original expression for the correlation coefficient in our matrix formalism and inserting the solution, the maximal correlation coefficient is given by

\beq
\rho^2_l = \frac{(\mathbf{c}\cdot \mathbf{k})^2}{\cl^{\kappa \kappa} \mathbf{c} \cdot  [\mathbf{Q}~\mathbf{c}] } =  \frac{\mathbf{k}\cdot  [\mathbf{Q}^{-1}~\mathbf{k}] }{\cl^{\kappa \kappa} } 
\eeq

\comment{As an aside, we note that $\mathbf{k}$ and $\mathbf{Q}$ can either be obtained directly by measuring the lensing-cross- and auto-correlations of the relevant fields $I_i$, or can be calculated analytically as follows (assuming we have sufficiently good models for the redshift kernels $W_i$):
\beq
k_i=\int \frac{dz H(z)}{ \eta^2(z)} W_i  W^\kappa(z) P(k = l/\eta(z),z) ,
\eeq
\beq
Q_{ij} = \int \frac{dz H(z)}{ \eta^2(z)} W_i W_j\left(P(k = l/\eta(z),z) + N(z) \right) 
\eeq}

\comment{Does the weighting of equation 37 make sense? We note that if the different fields $I_i$ are independent, we can use the definitions of $k_i$ and $Q_{ij}$ to find that the weighting reduces to
\beq
c_i I_i(\bl) \rightarrow \frac{\cl^{\kappa I_i}}{\cl^{I_i I_i}} I_i(\bl) = \frac{(\cl^{\kappa I_i})^2}{\cl^{I_i I_i} \cl^{\kappa \kappa}} \frac{\cl^{\kappa \kappa} I_i(\bl)}{\cl^{\kappa I_i}} =\rho^2_i \frac{\cl^{\kappa \kappa} I_i(\bl)}{\cl^{\kappa I_i}}.\eeq
This weighting is thus equivalent to: first, rescaling the field by ${\cl^{\kappa \kappa}}/{\cl^{\kappa I_i}}$; then, weighting the field by the $(S/N)^2$ of the cross-correlation (which is equal to the squared correlation coefficient $r^2$). The weighting thus appears sensible.}

Inspired by this result we rewrite the weights in terms of the correlation coefficients of two $I_i$ fields, $\rho_{ij}$, and the correlation coefficients of the $I_i$ fields with the true lensing convergence, $\rho_{i \kappa}$. We can rewrite the matrix $Q$ as (no sum)
\beq
Q_{ij} = \rho_{ij} \sqrt{\cl^{I_i I_i}  \cl^{I_j I_j} }; (Q^{-1})_{ij} =  (\rho^{-1})_{ij}/\sqrt{\cl^{I_i I_i}  \cl^{I_j I_j} }
\eeq
and
\beq
k_i = \rho_{i \kappa} \times \sqrt{ \cl^{I_i I_i} \cl^{\kappa \kappa}  }
\eeq

We obtain the following weight for a field $I_i$ (indices here are not summed over unless indicated):

\beq
c_i = \left[ \sum_j (\rho^{-1})_{ij} \rho_{ j \kappa }\times \sqrt{\frac{\cl^{\kappa \kappa}}{\cl^{I_i I_i}  }} \right] 
\eeq
so that

\beq
I = \sum_i  \left[ \sum_j (\rho^{-1})_{ij} \rho_{ j \kappa } \right]  \sqrt{\frac{\cl^{\kappa \kappa}}{\cl^{I_i I_i}  }} I_i
\eeq

The correlation coefficient for this optimal combination is
\beq
\rho^2 = \sum_{ij} \rho_{i \kappa} (\rho^{-1})_{ij} \rho_{ j \kappa }
\eeq

The power spectrum of this linear combination is
\beqn
\cl^{II} &=& \left( \sum_{iajb} \rho_{ia} \left[ (\rho^{-1})_{ij} \rho_{ j \kappa }\right] \left[ (\rho^{-1})_{ab} \rho_{ b \kappa }\right]   \right) \times\cl^{\kappa \kappa} \nonumber \\&=&\left[\sum_{ij }\rho_{i \kappa} (\rho^{-1})_{ij} \rho_{ j \kappa } \right] \times \cl^{\kappa \kappa} \eeqn
and the cross-spectrum with $\kappa$ takes the exact same value.
Hence
\beq
\cl^{II} = \cl^{\kappa I} = \rho^2 \cl^{\kappa \kappa} 
\eeq
We note that because $a =\cl^{\kappa I} / \cl^{I I} =1$, the linear combination is already correctly normalized (for our choice of $n=1$).

\comment{\subsection{Optimal weights for tracers with full redshift information}

For reference, we also include a calculation for a large-scale-structure (LSS) tracer $I$ with full redshift information. To construct the ideal delensing tracer, we must again maximize the (square of) the  correlation coefficient $\rho$ (which is equivalent to cross-power signal-to-noise):

\begin{equation}
\rho^2_l = \frac{{\cl^{\kappa I}}^2}{\cl^{\kappa \kappa} \cl^{II}}
\end{equation}

Here we use standard expressions for the cross-powers of the lensing convergence and the large-scale structure tracer
\beq
C_l^{\kappa I} = \int \frac{dz H(z)}{ \eta^2(z)} W^\kappa(z) W^I(z) P(k = l/\eta(z),z)
\eeq
as well as the auto powers
\[
C_l^{\kappa \kappa} = \int \frac{dz H(z)}{ \eta^2(z)} \left[ W^\kappa(z)\right]^2P(k = l/\eta(z),z)
\]
\beq
C_l^{I I} = \int \frac{dz H(z)}{ \eta^2(z)} \left[ W^I(z) \right]^2 \left(P(k = l/\eta(z),z) + N(z) \right).
\eeq

The $W$ functions are the relevant redshift kernels for lensing
\beq
W^\kappa(z) = \frac{3}{2 H(z)} \Omega_0 H_{0}^{2} (1+z) \eta(z) \frac{\left(\eta^{LS}-\eta(z)\right)}{\eta^{LS}}
\eeq
and the LSS tracer $I$
\beq
W^I(z) = \frac{b(z)\frac{dI}{dz}}{ \left[\int dz'\frac{dI}{dz'}\right]} 
\eeq

respectively, where $\eta(z)$ is the comoving distance to redshift $z$, $\hat{\mathbf{n}}$ is a direction on the sky, $\eta^{LS}$ is the comoving distance to the last scattering surface, $z_{LS}$ is the redshift of the last scattering surface, $H(z)$ is the Hubble parameter, $\Omega_0$  and $H_{0}$ represent the present values of the matter density parameter and the Hubble parameter, and $b$ and $\frac{dI}{dz}$ are the LSS tracer bias and redshift distribution.

We will assume we can weight redshifts and scales differently to construct a delensing tracer map; we hence multiply the kernel $W^I$ by a relevant weighting $f(z,l)$, such that $W^I \rightarrow f ~W^I $. We now seek to find $f$ such that it maximizes the cross-correlation coefficient. 

This cross-correlation coefficient becomes
\beq
\rho^2_l = \frac{  \left[ \int \frac{dz H(z)}{ \eta^2(z)} f(z,l) W^\kappa(z) W^I(z) P(k = l/\eta(z),z) \right]^2 }{\cl^{\kappa \kappa}  \int \frac{dz H(z)}{ \eta^2(z)} f^2(z,l) \left[ W^I(z) \right]^2 \left(P(k = l/\eta(z),z) + N(z) \right) }
\eeq

We now maximize $\ln{\rho^2}$ with respect to $f$
\beqn
0&=&\frac{\delta}{\delta f} \ln(\rho^2) =
\frac{  2  \int \frac{dz H(z)}{ \eta^2(z)} W^\kappa(z) W^I(z) P(k = l/\eta(z),z)  }{   \int \frac{dz H(z)}{ \eta^2(z)} f(z,l) W^\kappa(z) W^I(z) P(k = l/\eta(z),z)   } \nonumber \\&-& \frac{ 2  \int \frac{dz H(z)}{ \eta^2(z)} f(z,l) \left[ W^I(z) \right]^2 \left(P(k = l/\eta(z),z) + N(z) \right)    }{  \int \frac{dz H(z)}{ \eta^2(z)} f^2(z,l) \left[ W^I(z) \right]^2 \left(P(k = l/\eta(z),z) + N(z) \right)   }
\eeqn
It can be seen that this equation is solved if
\beq
f W^I (P+N) = W^\kappa P
\eeq
and so our weight function which maximizes the cross-correlation coefficient is
\beq
f(z,l) = \frac{P(z,l)}{P(z,l)+N(z,l)}\times W^\kappa(z)/W^I(z)
\eeq
which gives a maximal correlation coefficient of
\beq
\rho^2_l = \frac{ \int \frac{dz H(z)}{ \eta^2(z)} \left[ W^\kappa(z)\right]^2  \frac{P(z,l)}{P(z,l)+N(z,l)}  P(k = l/\eta(z),z) }{\cl^{\kappa \kappa}  }
\eeq}

\end{document}